\documentstyle[12pt,preprint,aps]{revtex}%
\begin{document}
\title
{Spin holography II}

\author
{Eddy Timmermans}
\address{Institute for Theoretical Atomic and Molecular Physics}
\address{Harvard-Smithsonian Center for Astrophysics}
\address{Cambridge, MA 02138}

\author
{G. T. Trammell and J. P. Hannon}
\address{Physics Department, Rice University}
\address{Houston, TX 77251-1892}

\date{\today}
\maketitle
\begin{abstract}

The interference of the directly emitted
photoelectron wave and the wave scattered coherently by neighboring atoms
gives holographic fringes in the photoelectron
emission intensity
$I(\hat{\bf R})$.  
In the electron emission holography technique in surface physics, 
$I(\hat{\bf R})$ is inverted holographically to give
a 3D-image of the 
environment of the source atom.  
Earlier \cite{usl}--\cite{us2}, we pointed out that the polarization
pattern ${\bf P}(\hat{\bf R})$ similarly can be viewed as a hologram of
the spin environment of the source atom by virtue of the exchange scattering
of the photoelectron by the neighboring atoms.
In this paper, we point out that spin-orbit correlations
in the photoelectron initial state are responsible for
holographic spin-dependent contributions to the intensity hologram 
$I(\hat{\bf R})$, even if the directly emitted photoelectrons 
are unpolarized.  This remarkable result implies that the emission intensity
contains spin information just as the polarization pattern
${\bf P}(\hat{\bf R})$.  Although the spin dependent signal in the
hologram is rather small ($\sim 5 \% $
in most cases of interest), 
we show how spin information can be extracted from the
intensity hologram, making use of the point symmetry of the 
environment of the source atom.
This way of analyzing photoelectron intensity holograms to extract
short-range spin information opens up a new avenue for surface
magnetism studies.

\end{abstract}


\narrowtext

\section{Introduction}

A photoelectron emitted from an atom may be scattered by neighboring atoms.
The scattered waves interfere with the unscattered wave and the intensity 
pattern at large distances, $I(\hat{\bf R})$, forms a hologram which can be 
analyzed to yield an image of the neighborhood of the electron emitting atom
\cite{Bart}--\cite{Szoke}.

	If the neighboring atoms have spins, then exchange scattering results 
in the polarization of the outgoing electrons and in addition to the 
intensity $I(\hat{\bf R})$, the polarization pattern ${\bf P}(\hat{\bf R})$
also forms a hologram which can be analyzed to yield an image of the spins
of the neighboring atoms.  While this scheme of `spin holography' is
elegant, it suffers from the fact that present day electron polarimeters
are very inefficient (eff. $\sim 10^{-4}$) detracting from the practical
utility of this way of imaging the spins.  Therefeore, the
existence of an alternate method to extract the
spin information from the intensity hologram $I(\hat{\bf R})$,
such as the scheme that we discuss in this paper, is of interest.

	Spin holography is a holography scheme yielding spin information 
on the near-neighborhood 
probed by electron holography.  This is in contrast to the
long range order and bulk magnetism determinations in magnetic 
diffraction methods using neutrons or X-rays.	
Since the effect of exchange interaction on the electron-atom
scattering becomes less pronounced as the kinetic energy of the electron 
increases, it is likely that spin holography will prove to be most useful if
the photoelectron has relatively low energy ($\sim$ 100-200 eV).
In this energy range, the mean-free path of the electron is of the
order of a few nanometers and only atoms within that distance 
of the photoelectron source atom will
contribute significantly to the hologram.
Therefore, we expect that electron holographic spin determination
will be useful as a probe of surface magnetism, with the source atoms
lying within the first few surface layers or as adatoms adhering to
the surface.  As an example where holographic spin determination
is useful, we can use adatom sources on the surface of a ferromagnet
to determine the spins of the first few layers, while the surface 
information is difficult to obtain from `diffraction' measurements.

	Consider the example of a substrate with the first two layers 
magnetized, layer $\# 1$ having all its atoms with spin $s_{a}$ and layer $\# 2$
having spin $s_{b}$.  Then while the holographic method with the sources,
e.g. adatoms on the top surface, would yield a proper image of the 
neighboring atoms on the two layers, diffraction measurements can only 
distinguish the actual magnetic surface ($1_{a}$, $2_{b}$)
from the surface with the spins of the layers reversed ($1_{b}$, $2_{a}$),
by detailed compairison of the diffraction data with numerical multiple 
scattering calculations.

Clearly, if the photoelectrons are $\it{polarized}$ then the
$\it{intensity}$ holograms, $I(\hat{\bf R})$, are sensitive to the spins 
of the neighboring scatterers.  However, a new and important feature
of the `spherical wave holograms' is that the photoelectrons need not be 
polarized for $I(\hat{\bf R})$ to be used to image the spins of the 
neighboring atoms: unpolarized electrons suffice if there are spin-orbit
correlations.  As a consequence, if the photoelectrons are ejected from
inner core spin-orbit split subshells (e.g. $L_{2}$ or $L_{3}$),
$\it{even \; by \; linearly \; polarized \; or \; unpolarized \;
incident \; light}$, then as we show below, their intensity holograms
$I(\hat{\bf R})$, can be analyzed to image the spins.  This is an 
interesting and important new result.  It is important that linearly polarized
synchrotron radiation can be used to image the spins.  It is interesting for the
same reason -- one is surprised (at least at first) that linearly polarized
light ejecting electrons from a complete inner atomic subshell gives an
intensity pattern with a linear dependence on the spins ${\bf s}_{i}$
of the neighboring atoms.  (Indeed, the analysis of these matters lead
us to interesting new spinor calculus results \cite{usp}).  Below, we discuss
the cases where linearly polarized, circularly polarized and unpolarized 
light is used to eject the photoelectrons.

	The paper is organized as follows.  In section II, we calculate
the photoelectron hologram, including the exchange scattering contributions
which give a linear dependence on the spins of the atoms in the 
near-neighborhood of the photoelectron source atom.  
Often the emitting atoms are centers of point symmetry, e.g. $C_{nv}$, for
the crystalline surfaces being investigated.  In those cases, we can use 
symmetry to aid in spin determination, as we discuss in section III.
We then proceed in section IV to illustrate the technique
for a specific example of a  $C_{4v}$ environment.
In section V, we discuss various schemes to extract spin information from
the holograms.  Finally, in section VI, we comment on the interesting case
of rare earth magnetism, for which the theory of spin holography, as
formulated below, needs to be generalized. 
We conclude and summarize the results in section VII.

\section{Electron Holograms}

        Initially, to be definite, we consider photoelectrons emitted from a 
$p_{1/2}$
subshell of a source atom following the absorption of photons linearly polarized
in the z-direction.  The $p_{1/2}$ subshell 
consists of two electrons with magnetic quantum numbers
$m_{j}$ = $+1/2$ and $m_{j} = -1/2$ respectively.  
The primary electron wave emitted
from the $m_{j}$-state is a spherically outgoing wave of wave number k,
\begin{equation}
|\psi^{0}_{m_{j}} \rangle 
\sim |\chi_{m_{j}} (\hat{\bf r}) \rangle \; \frac{\exp(ikr)}{r}  \; \; , 
\label{e:pw}
\end{equation}
where $|\chi_{m_{j}} (\hat{\bf r}) \rangle$ 
is the appropriate angular dependent 
spinor of the primary electron wave.
In the reference frame pictured in Fig.(1), these spinors (in the dipole
approximation) are proportional to
\begin{eqnarray}
&&|\chi_{1/2} (\hat{\bf R}) \rangle =
\left( 
\begin{array}{c}
\cos^{2} (\theta ) + c' \\
\sin (\theta ) \cos(\theta ) \exp (i\phi )
\end{array} \right) \; \; ;
\nonumber \\
\nonumber \\
&&|\chi_{-1/2} (\hat{\bf R}) \rangle =
\left(
\begin{array}{c}
\sin (\theta ) \cos(\theta ) \exp (-i\phi ) \\
\cos^{2} (\theta ) + c' \\
\end{array} \right) \; \; , 
\label{e:sp}
\end{eqnarray}
where $\theta$ is the usual polar angle and $\phi$ the azimuthal angle of 
$\hat{\bf R}$ in Eq.(\ref{e:sp}), and
in the following,
$c'=(c-1)/3$, where c is the ratio of the radial matrix elements
$M_{0}$ ($M_{2}$) and corresponding phase shifts $\delta_{0}$ ($\delta_{2}$)
for emission into the continuum s- or d-states, c = $(M_{0}/M_{2}) 
\exp[i(\delta_{0}-\delta_{2})]$.

        The primary wave is scattered by an atom i, giving a scattered wave 
$|\psi^{i}_{m_{j}} \rangle$.  We suppose that $\rm{k \; r}_{i} >> 1$ and
approximate the scattered wave by that of an incident
plane wave with wave vector $k\hat{\bf r}_{i}$, see \cite{Bart2}--\cite{Bart4}
for a discussion of 
this approximation
in holography.  In this case  $|\psi^{i}_{m_{j}} \rangle$ takes the form,
\begin{eqnarray}
|\psi^{i}_{m_{j}} (\hat{\bf R}) \rangle
= \frac{\exp (ik r_{i}) } { r_{i} } \;
f(\hat{\bf R},\hat{\bf r}_{i}) \; \exp (-ik\hat{\bf R}\cdot \hat{\bf r}_{i})
\; |\chi_{m_{j}} (\hat{\bf r}_{i}) \rangle \; \frac{\exp(ikR)}{R} \; \; ,
\label{e:psimj}
\end{eqnarray}
where $f(\hat{\bf R},\hat{\bf r}_{i})$ is the scattering amplitude
for the scattering of the photoelectron, incident along the 
$\hat{\bf r}_{i}$-direction and scattered in the $\hat{\bf R}$-direction.

	If now we neglect multiple scattering, then  
$|\psi_{m_{j}} ({\bf r}) \rangle = |\psi^{0}_{m_{j}} ({\bf r}) \rangle
+ \sum_{i} |\psi^{i}_{m_{j}} ({\bf r}) \rangle$
represents the photoelectrons emitted from the $m_{j}$-state.  
The number of electrons
emitted in direction $\hat{\bf R}$, per unit of solid angle is, 
up to a constant of proportionality
which depends on the intensity of the incident photons and the counting time,
equal to the `intensity' 
$I(\hat{\bf R}) \equiv R^{2} \; \sum_{m_{j}}  \langle \psi_{m_{j}} ({\bf R}) |
\psi_{m_{j}} ({\bf R}) \rangle$.  Using the above expression we find
\begin{eqnarray}
I(\hat{\bf R}) &=& \sum_{m_{j}} \; \left[
\langle \chi_{m_{j}} (\hat{\bf R}) | \chi_{m_{j}} (\hat{\bf R}) \rangle +
\; \sum_{i}  2 \; Re \left( [\exp(ikr_{i})/r_{i}]
\right.
\right.
\nonumber \\
&& 
\left.
\left.
\langle \chi_{m_{j}} (\hat{\bf R}) | f(\hat{\bf R},\hat{\bf r}_{i}) |
\chi_{m_{j}} (\hat{\bf r}_{i}) \rangle \exp (-ik \hat{\bf R} \cdot 
\hat{\bf r}_{i}) \right) + \cdots \right] 
\label{e:i}
\end{eqnarray}
In (\ref{e:i}) we keep only zero and first order terms in the 
scattering amplitudes, f,
neglecting the effects of multiple scattering and self interference terms (such
terms tend to degrade the holographic images and can be partially eliminated by
using special kernels in the hologrgaphic transform, as well as sums over
holograms collected at different energies, as has been discussed extensively 
in the literature \cite{Bart5}--\cite{Ton}).

	The phase factor $\exp(-ik\hat{\bf R}\cdot {\bf r}_{i})$
in Eq.(\ref{e:i}) gives holographic fringes in the 
angular intensity pattern; and transforms of $I(\hat{\bf R})$ 
by suitable kernels proportional to $\exp(ik\hat{\bf R}\cdot {\bf r})$, where 
${\bf r}$ is the parameter of the transform, peak
in the vicinities of ${\bf r} = {\bf r}_{i}$, yielding atomic images.

        If scattering atom i has a spin, then the scattering 
amplitude for coherent scattering,
$f(\hat{\bf R},\hat{\bf r}_{i})$, depends on the 
thermally averaged expectation value of the spin 
vector, ${\bf s}_{i}$, of the scattering atom,
\begin{eqnarray}
f(\hat{\bf R},\hat{\bf r}_{i}) =  f_{0}(\hat{\bf R},\hat{\bf r}_{i}) + 
f_{s}(\hat{\bf R},\hat{\bf r}_{i}) \sigma \cdot {\bf s}_{i} \; \; ,  
\label{e:f}
\end{eqnarray}
where the first term is the spin independent contribution $f_{0}$
and the second term is the exchange contribution, proportional to
the scalar product of ${\bf s}_{i}$ with the 
Pauli-spin operator $\sigma$ of the photoelectron.   
If we could treat the scattering atoms as spherically symmetric
then $f_{0}$ and $f_{s}$ would only depend on $\hat{\bf R} \cdot 
\hat{\bf r}_{i}$.  In reality, the valence shell electron 
distributions are affected
by the presence of the neighboring atoms.  The charge (and spin) distribution
within a Wigner-Seitz cell representing an atom is not spherically symmetric
about the cell's center.  This is particularly true for for the valence shell
electrons having the uncompensated spins which is our primary concern.
In the copper oxides of interest in the high $T_{c}$ materials, for
example, the uncompensated spins are believed to occupy $x^{2}-y^{2}$ 
d-orbitals on the copper ions.  
Furthermore, we note that
the scattering amplitude (\ref{e:f}) describes exchange scattering by
a magnetic ion if the orbital moments are `quenched' and the spin
direction is free to change without changing the spatial distribution
of the charge or spin distributions.
This description is accurate for transition elements, but fails for
rare earth ions (except for ${\rm Gd}^{3+}$, in which case $L = 0$).
In that case, there is strong
spin-orbit correlation and both $f_{0}$ and $f_{s}$ in Eq.(\ref{e:f})
must themselves be taken to be functions of ${\bf s}_{i}$.  
In the following, we shall assume a scattering amplitude of the form
(\ref{e:f}) and we will
return to the more complicated but important case of rare earth ions 
later (section VI).
In any case, due to the spin dependence of the scattering amplitude,
it should be evaluated inside the spinor bracket of Eq.(\ref{e:i}).  With 
the scattering amplitude (\ref{e:f}) we obtain
\begin{eqnarray}
I(\hat{\bf R}) &=& \langle \langle \hat{\bf R} | \hat{\bf R} \rangle \rangle
+ 2 Re \left( \sum_{i} \; \; [\exp(ikr_{i})/r_{i}]  \; \times
\right.
\nonumber \\
&&
\left.
\left[
f_{0}(\hat{\bf R},\hat{\bf r}_{i}) \; 
\langle \langle \hat{\bf R} | \hat{\bf r}_{i} \rangle \rangle
+ f_{s}(\hat{\bf R},\hat{\bf r}_{i}) \; 
\langle \langle \hat{\bf R} | \sigma | \hat{\bf r}_{i} \rangle \rangle
\cdot {\bf s}_{i} \; \right] \exp (-ik{\bf R} \cdot {\bf r}_{i}) + \ldots
\; \right) \; ,
\label{e:i2}
\end{eqnarray}
where we represent the sum over the initial electron states $m_{j}$ 
of the `spinor interference brackets' by
\begin{eqnarray}
&& \langle \langle \hat{\bf R}| \hat{\bf r}_{i} \rangle \rangle
\; \; = \sum_{m_{j}} \; \langle \chi_{m_{j}} (\hat{\bf R}) | \chi_{m_{j}}
(\hat{\bf r}_{i}) \rangle  \; \; \; ,
\nonumber \\
&& \langle \langle \hat{\bf R} | \sigma | \hat{\bf r}_{i} \rangle \rangle
= \sum_{m_{j}} \; \langle \chi_{m_{j}}(\hat{\bf R}) | \sigma |
\chi_{m_{j}} (\hat{\bf r}_{i}) \rangle \; . 
\label{e:db}
\end{eqnarray}

	Eq.(\ref{e:i2}) includes an interference term that is proportional 
to $f_{s}$ and the
spin ${\bf s}_{i}$ of the scattering atom so that the intensity is 
sensitive to the spins of
the neighboring atoms if the
`spin interference matrix element', $\langle \langle \hat{\bf R} | \sigma
| \hat{\bf r}_{i} \rangle \rangle$, does not vanish.
Generally, even for linearly polarized light (or indeed,
unpolarized light) the spin interference matrix element 
$\it{does \; not \; vanish}$  if the
electron is ejected from a spin-orbit split sublevel.

        In our example of photoemission from a $p_{1/2}$ level, 
the interference 
matrix elements for photons of arbitrary polarization
$\hat{\epsilon}$ ($\hat{\epsilon} = \hat{\bf z}$, for z-linearly
polarized, $\hat{\epsilon} = (\hat{\bf x} \pm i \hat{\bf y})/\sqrt{2}$
for right- and left-hand polarized light etc ....) are equal to
\begin{eqnarray}
&& \langle \langle \hat{\bf R} | \hat{\bf r} \rangle \rangle
= 2 \{ (\hat{\epsilon}^{*} \cdot \hat{\bf R}) \; (\hat{\epsilon} \cdot
\hat{\bf r} ) \; (\hat{\bf r} \cdot \hat{\bf R}) \; + |c'|^{2}
+ c'^{*} \; (\hat{\epsilon} \cdot \hat{\bf r}) \; 
(\hat{\epsilon}^{*} \cdot \hat{\bf r})
\; + c' (\hat{\epsilon}^{*} \cdot \hat{\bf R}) \;
(\hat{\epsilon} \cdot \hat{\bf R}) \; \} \; ,
\nonumber \\
&&
\langle \langle \hat{\bf R} | \sigma | \hat{\bf r} \rangle \rangle
= -2 \; i \; \{ 
|c'|^{2} \; (\hat{\epsilon}^{*} \times \hat{\epsilon}) - c' \; 
(\hat{\epsilon} \times \hat{\bf R}) \; (\hat{\epsilon}^{*} \cdot
\hat{\bf R}) + c'^{*} \; (\hat{\epsilon}^{*} \times \hat{\bf r}) \;
(\hat{\epsilon} \cdot \hat{\bf r}) 
\nonumber \\
&& \; \; \; \; \; \; \; \; \; \; \; \; \; \; \;
\; \; \; \; \; \; \; \; \; \; \; \; \; \; \;
+ (\hat{\epsilon}^{*} \cdot \hat{\bf R}) \; 
(\hat{\epsilon} \cdot \hat{\bf r}) \; (\hat{\bf R} \times \hat{\bf r}) 
\} \; \; . 
\label{e:int}
\end{eqnarray}

	For emission from a filled $p_{3/2}$-shell the intensity 
interference matrix element,
$\langle \langle \hat{\bf R} | \hat{\bf r} \rangle \rangle$,
is twice the intensity matrix element of Eq.(\ref{e:int}), as a
consequence of the $p_{3/2}$-shell having twice as many electrons.
Similarly, the $p_{3/2}$ spin interference matrix element,
$\langle \langle \hat{\bf R} |\sigma | \hat{\bf r} \rangle \rangle$
is the negative of the spin interference matrix element shown on the
last line of Eq.(\ref{e:int}) \cite{spinorb},
which follows from the fact that the $p_{3/2}$ and $p_{1/2}$ spin matrix
elements add up to the spin matrix element for photoemission from a
p-shell which, in the absence of spin-orbit interaction, vanishes.

	It is a feature of spherical wave electron holography that the
interference between electron waves emitted in $\it{different}$ directions
is recorded.  As a consequence, it is not the spin polarization
of the electrons emitted in the direction of the scattering atom 
$\hat{\bf r}_{i}$ that gives the holographic spin dependence, but the 
interference matrix element of the emission directions $\hat{\bf R}$ and
$\hat{\bf r}_{i}$, $\langle \langle \hat{\bf R} | \sigma | \hat{\bf r}_{i}
\rangle \rangle$.  The occurrence of these quantities in our theory leads
to interesting new aspects of angular momentum theory.

	That the spin matrix element is of the form of Eq.(\ref{e:int})
can be understood from general considerations -- the matrix element must
be `bilinear' in the photon polarization, meaning that each term contains
$\hat{\epsilon}$ and $\hat{\epsilon}^{*}$.  Furthermore, the spin matrix
element must be a pseudo vector made up of $\hat{\epsilon}$,
$\hat{\epsilon}^{*}$, $\hat{\bf R}$, and $\hat{\bf r}_{i}$.  In the dipole
approximation, the wave emitted from a p-shell is an admixture of an s and
a d-wave, from which follows that each term contains $\hat{\bf R}$ and
$\hat{\bf r}_{i}$ twice or not at all.  Based on these considerations we
can deduce the general form of the individual terms that make up the 
$p_{1/2}$ (or $p_{3/2}$) spin matrix element.

	Some remarkable results follow from Eq.(\ref{e:int}). For example,
the spin interference matrix element for z-linearly polarized light does not 
vanish :
\begin{equation}
\langle \langle \hat{\bf R} | \sigma | \hat{\bf r}_{i} \rangle 
\rangle_{\bf z} =
\; -2 \; i \; \{ - c' (\hat{\bf z} \times \hat{\bf R}) \; (\hat{\bf z} \cdot
\hat{\bf R}) \; + c'^{*} (\hat{\bf z} \times \hat{\bf r}_{i} )
\; (\hat{\bf z} \cdot \hat{\bf r}_{i}) 
+ (\hat{\bf z} \cdot \hat{\bf R}) \; (\hat{\bf z} \cdot \hat{\bf r}_{i})
\; (\hat{\bf R} \times \hat{\bf r}_{i}) \}   \; ,
\label{e:spinlin}
\end{equation}
implying that linearly polarized photons give spin dependent photoelectron
intensities.  The expected spin of the directly emitted photoelectrons in
the ${\bf R}$-direction is proportional to $\langle \langle
\hat{\bf R} | \sigma | \hat{\bf R} \rangle \rangle$, which can be obtained
from Eq.(\ref{e:spinlin}), putting $\hat{\bf r}_{i} \rightarrow \hat{\bf R}$.
This gives an electron spin polarization that is proportional to
\begin{equation}
\langle \langle \hat{\bf R} | \sigma | \hat{\bf R} \rangle \rangle_{\bf z}  =
- \; \frac{4}{3} \; \sin (\delta_{0}-\delta_{2}) \; \frac{M_{0}}{M_{2}} \;
(\hat{\bf z} \times \hat{\bf R}) \; (\hat{\bf z} \cdot \hat{\bf R}) \; \; ,
\label{e:spind}
\end{equation}
where we replaced the imaginary part of the $c'$-parameter by its expression
in terms of the phase shifts and the radial dipole matrix elements.  The
result of Eq.(\ref{e:spind}) makes it clear that the resulting spin polarization
is in fact caused by the interference of the outgoing s and d-waves, as was 
pointed out previously \cite{Heinz}.
Nevertheless, even if the spin polarization (\ref{e:spind}) of the 
unscattered photoelectrons vanish, because, for example, the electron is 
emitted as a pure d-wave, the interference spin matrix element does 
$\it{not}$ vanish. The pure d-wave photoemission corresponds to 
$c'=-1/3$ giving
\begin{equation}
\langle \langle \hat{\bf R} | \sigma | \hat{\bf r}_{i} \rangle \rangle _{\bf z}
= -2 \; i \; \{ \frac{1}{3}
(\hat{\bf z} \times \hat{\bf R}) \; (\hat{\bf z} \cdot
\hat{\bf R}) - \frac{1}{3} (\hat{\bf z} \times \hat{\bf r}_{i}) \;
(\hat{\bf z} \cdot  \hat{\bf r}_{i}) \;
+ \; (\hat{\bf z} \cdot \hat{\bf R}) \; (\hat{\bf z} \cdot \hat{\bf r}_{i})
\; (\hat{\bf R} \times \hat{\bf r}_{i} ) \} \; \; .
\label{e:spindw}
\end{equation}

	Thus the d-wave photoelectron incident upon and scattered from the 
scattering atom is not spin polarized, 
(i.e. $\langle \langle \hat{{\bf r}}_{i}|\sigma|
\hat{{\bf r}}_{i} \rangle \rangle = 0$),
nor is the unscattered wave emitted
in the  $\hat{\bf R}$-direction polarized,
(i.e. $\langle \langle \hat{{\bf R}}|\sigma|\hat{{\bf R}} 
\rangle \rangle = 0$); 
but the interference between the photoelectron waves emitted in the
${\bf r}_{i}$-direction and scattered into the
direction $\hat{\bf R}$ with the waves emitted directly in that direction
does result in a polarization and gives a photoelectron intensity
sensitive to the spin of the scattering atom.  Another interesting
consequence of this
angular correlation in the photoelectron spin is that the interference
of the directly emitted and scattered electron waves gives a finite spin
polarization to the photoelectrons, even though the directly emitted
electrons are spin unpolarized and the scattering atom is not magnetic.

	Perhaps even more surprising than the fact that incident linearly
polarized light gives rise to spin-sensitive holograms is that so does
unpolarized light.  Note that unpolarized light incident along the z-direction,
say, on the source atom gives photoelectron intensities equal to the
sum of those for x- and y-linearly polarized beams.  Now the sum of 
x-,y- and z-polarized beams gives
\begin{equation}
\langle \langle \hat{\bf R} | \sigma | \hat{\bf r}_{i} \rangle \rangle 
= - \; 2 \; i \; (\hat{\bf R} \times \hat{\bf r}_{i} ) \; 
(\hat{\bf r}_{i} \cdot \hat{\bf R}) \; \; ,
\label{e:xyz}
\end{equation}
thus,
\begin{equation}
\langle \langle \hat{\bf R} | \sigma | \hat{\bf r}_{i} \rangle \rangle
_{\bf x} \; + 
\langle \langle \hat{\bf R} | \sigma | \hat{\bf r}_{i} \rangle \rangle
_{\bf y} = 
- \langle \langle \hat{\bf R} | \sigma | \hat{\bf r}_{i} \rangle \rangle
_{\bf z} - 2 \; i \; (\hat{\bf R} \times \hat{\bf r}_{i}) \;
({\bf r}_{i} \cdot \hat{\bf R}) \; \; ,
\label{e:z}
\end{equation}
and substituting from Eq.(\ref{e:spinlin}) we see that unpolarized light also
gives a non-vanishing spin contribution to the photoelectron intensity and thus
can serve to photoemit electrons with an intensity
hologram that is sensitive to the spins of the nearby atoms.

	These surprising results are caused by the spin-orbit correlation in 
the initial shell, which is preserved in the emitted d-waves.  The remarkable 
consequences clearly indicate that the interference matrix elements contain
novel and interesting physics.

	Returning to the intensity hologram of Eq.(\ref{e:i2}),
it is clear from the subsequent discussion of the interference matrix 
elements that the spin contributions to the intensity hologram
don't vanish if the electrons are
emitted from an atomic spin-orbit split shell.  
The spin contributions are, however,
relatively small, only making an addition of the order of
$|f_{s} {\bf s}_{i} / f_{0}|$ ($\simeq 0.05$, e.g.,
for electron scattering by an iron atom at 100 eV) to the ordinary, spin
independent `charge' hologram.  {\it The  challenge  then  is  to
 extract  the  spin  information  from  the  hologram}.

\section{Spin Analysis}

	The question becomes how to isolate the spin dependent part of the 
hologram from the much larger `charge' part background.

	Referring back to Eq.(\ref{e:i2}), if we were able to reverse all the
${\bf s}_{i}$'s of the sample then the difference hologram
would contain only the spin dependent terms.  Reversing the sample's spins,
however, would be feasible only for ferro- or ferrimagnetic samples.
Alternatively, the difference of holograms below and above the magnetic
ordering temperatures can serve to determine the spins in the cases where we
can assume that the change of the spin independent factors in Eq.(\ref{e:i2})
is small.

	Other than those two methods depending on changing the magnetization
of the sample by changing the magnetic field or the temperature, which may not
be feasible, there are other methods which can serve our purposes:

	If the primary wave, (\ref{e:sp}), had only an s-wave component
rather than the d-s mixture, then only the $|c'|^{2}$ terms in Eq.(\ref{e:int})
would be present.  In that case $\langle \langle \hat{\bf R} | \hat{\bf r}_{i}
\rangle \rangle \sim |c'|^{2}$, $\langle \langle \hat{\bf R} | \sigma
| \hat{\bf r}_{i} \rangle \rangle _{\pm} \sim \pm |c'|^{2} \; \hat{\bf z}$
for $\hat{\epsilon} = (\hat{\bf x} \pm \hat{\bf y})/\sqrt{2}$, and
the difference holograms for incident right and left-hand circularly
polarized light around, successively x,y and z-axes would determine 
all three components of the spin vectors.
In fact, however, the s-wave component is almost an order of magnitude smaller 
than the dominant d-wave contribution in the case of photoemission from
a p-shell at the relevant emission energies ($\sim 100 - 200$ eV).
Furthermore, we note that
the extraction of the spin terms from right-left
polarization differences contain spin independent `charge' contributions.
Indeed,
with Eq.(\ref{e:int}) we find
\begin{equation}
\langle \langle \hat{\bf R} | \hat{\bf r}_{i} \rangle \rangle _{+}
\; - \langle \langle \hat{\bf R} | \hat{\bf r}_{i} \rangle \rangle _{-}
\; = - \; 2 \; i \; \left[
 (\hat{\bf r}_{i} \cdot \hat{\bf R} ) \;
(\hat{\bf R} \times \hat{\bf r}_{i}) \cdot \hat{\bf z}
\;  \right] \; \; ,
\label{e:intpm}
\end{equation}
and referring back to Eq.(\ref{e:i2}), we see that the difference
hologram does indeed contain `charge' contributions.
	
	We can, however, use $\it{symmetry}$ to aid in spin determination.
Often the emitting atoms are centers of point symmetry, e.g. $C_{nv}$, for the
crystalline surfaces being investigated.  Any differences detected
after subjecting the hologram to a point symmetry group operation is then
the result of the symmetry breaking spin contributions.  Based on this
general idea, we work out practical schemes to construct `spin
holograms'.  For the sake of simplicity, we start by illustrating the
main idea for the case of reflection symmetry in subsection \ref{3a}.
In subsection \ref{3b}, we approach the problem from a more general perspective.

\subsection{Reflection Spin Holography}
\label{3a}

        Let $I(\hat{\bf R},\hat{\epsilon};m)$ represent the photoelectron
intensity emitted in the $\hat{\bf R}$-direction, for 
photon polarization $\hat{\epsilon}$, matter state m,
and let $I(\hat{\bf R}',\hat{\epsilon}';m')$
be the corresponding quantities following a spatial reflection.  Since
the Hamiltonian is invariant under reflections,
\begin{equation}
I(\hat{\bf R},\hat{\epsilon};m) = 
I(\hat{\bf R}',\hat{\epsilon}';m') \; \; . 
\label{e:ref}
\end{equation}
Also,
\begin{equation}
I(\hat{\bf R}',\hat{\epsilon}';m) = 
I(\hat{\bf R},\hat{\epsilon};m') \; \; ,
\label{e:ref2}
\end{equation}
since the square of the reflection is the unit operator.  Now `m' represents
the initial state of the source atom, centered at the origin, and the various 
neighboring atoms centered at ${\bf r}_{i}$, with spins ${\bf s}_{i}$, and
other properties, including valence electron spatial distributions, $o_{i}$ :
m = $[ {\bf r}_{1},{\bf s}_{1}, o_{1} ;  {\bf r}_{2},{\bf s}_{2}, o_{2} ,
\ldots , {\bf r}_{i},{\bf s}_{i}, o_{i} ; \ldots ]$, 
m' = $[ {\bf r}_{1}',{\bf s}_{1}', o_{1}' ; \ldots ]$.
Since ${\bf s}$ is an axial vector, ${\bf s}_{//}' = - {\bf s}_{//}$,
${\bf s}_{\perp}' = {\bf s}_{\perp}$, where the $//$ and $\perp$ subscripts
refer to components respectively parallel and perpendicular to the
plane of reflection.

        If the source atom is in a site with $C_{nv}$ point symmetry,
magnetic ordering of the neighboring atoms may break that symmetry, and
the difference hologram,
\begin{eqnarray}
I_{V} (\hat{\bf R},\hat{\epsilon};m) &\equiv& \frac{1}{2} \;
\left[ I(\hat{\bf R},\hat{\epsilon};m) - I(\hat{\bf R}',\hat{\epsilon}';m)
\right] \; ,
\nonumber \\
&=& \frac{1}{2} \; \left[ I(\hat{\bf R},\hat{\epsilon};m)
- I({\bf R},\hat{\epsilon};m') \right] \; \; ,
\label{e:iv}
\end{eqnarray}
contains only the symmetry breaking spin terms if the reflection plane,
V, is a symmetry plane.  This is so because if atom `j' is carried into atom 
`k' by the reflection then $o_{j}' = o_{k}$, by symmetry, and the 
non-spin dependent part of the hologram, the `charge' hologram is unchanged 
by the reflection.

	If atom `i' lies in the reflection plane then ${\bf r}_{i} =
{\bf r}_{i}' \;$,$ \; o_{i} = o_{i}' \;$,$ \; {\bf s}_{i,//} = -{\bf s}_{i,//}'
\;$,
$ \; {\bf s}_{i,\perp} = {\bf s}_{i,\perp}' \; $, and the contribution of `i'
to the difference hologram $I_{V}$ is (see Eq.(\ref{e:i2}))
\begin{equation}
I_{V}(\hat{\bf R}) = 2 \; Re \{ [\exp(ikr_{i})/r_{i}] \;
f_{s}({\bf R},{\bf r}_{i}) \langle \langle \hat{\bf R} | \sigma
| \hat{\bf r}_{i} \rangle \rangle \cdot {\bf s}_{i,//} \;
\exp(-ik\hat{\bf R}\cdot {\bf r}_{i}) \} \; \; .
\label{e:iv2}
\end{equation}

	On the other hand, if atom `j' does not lie in the reflection plane
`V' and is carried into atom `k' by the reflection then ${\bf s}_{k}
\rightarrow \frac{1}{2} ({\bf s}_{k} -{\bf s}_{j}')$ = $\frac{1}{2}({\bf s}_{k}
+ {\bf s}_{j,//} - {\bf s}_{j,\perp})$,
and the contribution of site `k' to $I_{V}$ becomes
\begin{equation}
I_{V}(\hat{\bf R}) = 2 \; Re \{ [\exp(ikr_{k})/r_{k}] \;
f_{s}({\bf R},{\bf r}_{k}) \langle \langle \hat{\bf R} | \sigma
| \hat{\bf r}_{k} \rangle \rangle \cdot \frac{1}{2} ({\bf s}_{k} 
+{\bf s}_{j,//}-{\bf s}_{j,\perp}) \;
\exp(-ik\hat{\bf R}\cdot {\bf r}_{k}) \} \; \; .
\label{e:iv3}
\end{equation}
The hologram $I_{V}$ given by summing the in-plane, Eq.(\ref{e:iv2}),
and out of plane, Eq.(\ref{e:iv3}), terms can then be transformed, as
we shall discuss later, to determine components of the spins of the various
atoms.

\subsection{Projection Spin Holography}
\label{3b}

        The above considerations regarding reflections illustrate the use of
symmetry to obtain spin holograms.  If G is an element, a reflection or a 
rotation,  of a point symmetry group which leaves the charge or chemical
neighborhood invariant, then the difference hologram involving G has only
the symmetry breaking spin contributions.  The argument is essentially that
given above.  For example, the difference between the emission intensity
$I(\hat{\bf R},\hat{\epsilon};m)$ measured in direction $\hat{\bf R}$ with
incident photons of polarization $\hat{\epsilon}$ and the intensity measured
in direction $G[\hat{\bf R}]$ with photons of polarization $G[\hat{\epsilon}]$,
$I(G[\hat{\bf R}],G[\hat{\epsilon}];m)$ contains only spin contributions
because the charge contributions cancel.

	To see that, we note that the Hamiltonian is invariant under all
of the transformations of the euclidean group (except for the small parity
non-conserving terms which are irrelevant for our considerations).
Thus, it follows that
\begin{equation}
I(\hat{\bf R},\hat{\epsilon};m) = I(G[\hat{\bf R}],G[\hat{\epsilon}];G[m]) \; ,
\label{e:genref}
\end{equation}
which is the generalization of (\ref{e:ref}), 
and where, as before, m represents the initial
state of the source atom and the neighboring atoms, 
m = $[{\bf r}_{1},{\bf s}_{1},o_{1};{\bf r}_{2},{\bf s}_{2};o_{2}, \ldots ]$.
Also, as in Eq.(\ref{e:ref2}),
\begin{equation}
I(G^{-1}[\hat{\bf R}],G^{-1}[\hat{\epsilon}];m) = 
I(\hat{\bf R},\hat{\epsilon} ; G[m]) \; \; ,
\label{e:genref2}
\end{equation}
where $G[m] = [G[{\bf r}_{1}],G[{\bf s}_{1}], G
[o_{1}]; \ldots , G[{\bf r}_{j}],[G[{\bf s}_{j}], G
[o_{j}]; \ldots ]$. Since $G$ and $G^{-1}$ belong to the symmetry
group, for every atom $j$ there is an atom $k$ for which ${\bf r}_{j} =
G[{\bf r}_{k}]$ (and $o_{j} = G[o_{k}]$).  If we denote
the label $k$ by $k = g[j]$, then
$G[{\bf s}_{k}] = G[{\bf s}_{g[j]}]$ and we can represent the 
transformed initial state as
$G[m] = [{\bf r}_{1},G[{\bf s}_{g[1]}], o_{1} ; \ldots
; {\bf r}_{j},G[{\bf s}_{g[j]}],o_{j} ; \ldots ],$ representing 
a neighborhood which has the same `charge' distribution' as m, but a different
spin arrangement: ${\bf s}_{1}$ apears now as $G[{\bf s}_{1}]$ at the
site of $G[{\bf r}_{1}]$ etc... .  Consequently, only spin
dependent terms do not cancel in the difference hologram
\begin{equation}
I_{G} \equiv \frac{1}{2}
\; \left[ I(\hat{\bf R},\hat{\epsilon};m)-I(G^{-1}[{\bf R}],
G^{-1}[\hat{\epsilon}];
m) \right] \; .
\label{e:ig}
\end{equation}
More generally, any linear combination of the type $\sum_{G} a(G) \;
I(G^{-1}[\hat{\bf R}],G^{-1}[\hat{\epsilon}];m)$, 
where the sum $\sum_{G}$ extends
over all elements G of the symmetry group of the `charge' environment and 
where the sum of the coefficients $a(G)$ vanishes,
$\sum_{G} a(G) = 0$, results in a 
hologram of which the charge contributions cancel out but not the spin terms.
In fact, the resulting angular pattern, 
$\sum_{G} a(G) \;
I(G[\hat{\bf R}],G[\hat{\epsilon}];m)$ is a hologram of the spin arrangment 
obtained from the actual spin distribution, by
replacing the spin ${\bf s}_{j}$ at site ${\bf r}_{j}$ 
by $\sum_{G} a(G) G[{\bf s}_{[g(j)]}]$ :
\begin{eqnarray}
I_{a}(\hat{\bf R},\hat{\epsilon};m) &\equiv&
\sum_{G} a(G) I(G^{-1}[\hat{\bf R}],G^{-1}[\hat{\epsilon}];m)
\; \; = \; \sum_{G} a(G) I(\hat{\bf R},\hat{\epsilon}; G[m])
\nonumber \\
&=& 2 \; \sum_{j} Re \{
[exp(ikr_{j})/r_{j}]\; f_{s}(\hat{\bf R},\hat{\bf r}_{j})
\; \exp(-ik\hat{\bf R} \cdot \hat{\bf r}_{j}) \;
\nonumber \\
&& \; \; \; \; \; \; \; \; \; \; \;
\langle \langle \hat{\bf R} | \sigma | \hat{\bf r}_{j} \rangle \rangle
\cdot \sum_{G} a(G) G[{\bf s}_{g[j]}] \} + \cdots \; \; .
\label{e:ia}
\end{eqnarray}

	Consider then the `star' of an atom at 
${\bf r}_{1}$, i.e., the atoms at $({\bf r}_{1}, {\bf r}_{2}, \cdots ,
{\bf r}_{N})$ into which ${\bf r}_{1}$ is carried by the operations of the
group.  (Thus for the group $C_{nv}$, N=n, if ${\bf r}_{1}$ lies in a
reflection plane, and N=2n, the order of the group, if it doesn't.)  Then
the spins on the N atoms constitute a 3N dimensional linear vector
space which transforms into itself under the group and which can be
decomposed into irreducible subspaces forming bases of the irreducible
representation of the group by the well known methods of group
theory \cite{LL}.  We may represent a vector in this 3N dimensional
space as ${\bf s}_{(N)} \equiv \{ {\bf s}_{1},{\bf s}_{2}, \ldots ; {\bf s}_{N}
\}$
= $ ( {\bf s}_{1x},{\bf s}_{1y},{\bf s}_{1z} ; {\bf s}_{2x},{\bf s}_{2y},
{\bf s}_{2z} ; \ldots )$.
From the general principles of group theory, we know that the 
${\bf s}_{(N)}$-vector can be decomposed into components that transform
according to the irreducible representation `$\alpha$' of the symmetry
group.  The decomposition, ${\bf s}_{(N)}$ = $\sum_{i,\alpha} 
{\bf s}_{(N)}^{(\alpha,i)}$, is achieved by means of 
the idempotent operator, $e^{\alpha}_{ii} = \frac{l_{\alpha}}{h} 
\sum_{G} D_{ii}^{\alpha} (G^{-1}) G$, 
which, when applied to a
linear vector space, selects out the $(\alpha,i)$-component of the vector :
\begin{equation}
{\bf s}_{(N)}^{(\alpha,i)} \; = \; \left( \frac{l_{\alpha}}{h} \right)
\; \sum_{G} D_{ii}^{\alpha}(G^{-1}) \; G[{\bf s}_{(N)}] \; \; ,
\label{e:proj}
\end{equation}
where $D_{ij}^{\alpha}(G)$ is the matrix representation of G in the
irreducible representation $\alpha$, $h$ is the order of the symmetry
group and i takes on values from 1 to $l_{\alpha}$, the dimension of the
representation.

	Returning to the spin hologram of Eq.(\ref{e:ia}), it is then clear
that identifying the $a(G)$ coefficients with $\left( \frac{l_{\alpha}}{h}
\right) \; D_{ii}^{\alpha} (G^{-1})$, gives a pattern that is the hologram
of the total spin ${\bf s}_{(N)}$ projected onto the irreducible
$(\alpha,i)$-mode ${\bf s}_{(N)}^{(\alpha,i)}$.

	This procedure, projecting out the irreducible modes of the 
photoelectron hologram,
\begin{equation}
I^{(\alpha,i)} (\hat{\bf R},\hat{\epsilon};m) 
= \left( \frac{l_{\alpha}}{h} \right) \; \sum_{G} \; D^{\alpha}_{ii} (G^{-1})
\; I(G^{-1}[\hat{\bf R}],G^{-1}[\hat{\epsilon}];m) \; \; ,
\label{e:ialpha}
\end{equation}
giving a hologram of the irreducible $(\alpha,i)$-mode of the spin
pattern, is completely general and gives the maximum information that
can be obtained from symmetry.

        Now we take $\alpha = 1$ for the identical representation,
$D^{1}(G) = 1$, so that the vector ${\bf s}^{1}_{(N)}$,
\begin{equation}
{\bf s}^{1}_{(N)} = \left(\frac{1}{l}\right)
\sum_{G} \; G^{-1}[{\bf s}_{(N)}] \; \; ,
\label{e:inv}
\end{equation}
is invariant under all G's and thus does not appear in the difference holograms
as in (\ref{e:ia}).  Such a vector is invariant under all symmetry operations G
and cannot be distinguished from the charge environment using symmetry alone.
The number of such linearly independent invariant vectors is given as usual
by the compound character averaged over all the group elements.  For
$C_{nv}$ there is one such vector for the N=n atom star:
${\bf s}_{i} = C \hat{\bf z} \times {\bf r}_{i}$.
Consequently, out of the
3n linearly independent vectors $3n-1$ can be determined from the holograms
(\ref{e:ialpha}). 
For the N=2n star there are three such vectors :
in addition to that for the N=n case, there are also $\pm \hat{\bf z}$, and
$\pm \hat{\bf r}_{i}$, where the signs for the two semi-stars, each consisting
of n atoms carried into each other by the 
$C_{n}$ subgroup, differ.  In this case,
one determines $6n-3$ linearly independent vectors by our method.

\section{An Example with $C_{4v}$ Symmetry}

        We illustrate the projection scheme for photoemission from an
atom placed in the environment pictured in Fig.(2).  Four identical 
neighboring atoms are located in the same horizontal plane with azimuthal 
angles $\phi = 0 \;$,$ \; \pi/2 \; $,$ \; \pi$ and $3\pi/2$, 
placed at equal distance from the emitting atom.  
The result is a `charge' environment of $C_{4v}$-symmetry.
        
	The $C_{4v}$-symmetry group has 8 elements (h=8): 
the identity operation E, 
2 rotations in the horizontal plane by $\pi/2$ : $\phi \rightarrow 
\phi + \pi/2$ ($C_{4}$) and $\phi \rightarrow \phi - \pi/2$ ($C_{4}'$),
one rotation by $\pi$ ($C_{2}$), and 4 reflections with respect to the 
vertical planes of azimuthal angle $\phi = 0 \;$ ($s_{V}'$), $\phi = \pi/4 \;$
($s_{d}$), $\phi = \pi/2 \;$ ($s_{V}$), $\phi = 3\pi/2 \;$ ($s_{d}'$).
There are 5 classes $\{ E, 2 C_{4}, C_{2}, 2 s_{V}, 2 s_{d} \}$ and 5
irreducible representations costumarily denoted by $\{ A_{1}, A_{2}, B_{1},
B_{2}, E \}$.  For the reader's convenience, we show the character
table of $C_{4v}$ in table I.
        
	The spin arrangement consists of four spin vectors (12 components) 
which generate a 12-dimensional representation of $C_{4v}$: $D^{(12)}$.  
Reducing the 12 components to modes that transform according to the
irreducible representations, we find one $A_{1}$, one $B_{1}$, two
$A_{2}$, two $B_{2}$ and three $E$-modes:
\begin{equation}
D^{(12)} \; = 1 \; A_{1} \oplus \; 1 \; B_{1} \; \oplus \; 2 \;
A_{2} \oplus \; 2 \; B_{2} \oplus 3 \; E \; .
\label{e:red}
\end{equation}
Each spin environment is a linear combination of these modes, pictured in 
Fig(3). Of the irreducible modes, only the $A_{1}$-mode, invariant under all
$C_{4v}$-operations cannot be determined by means of the projection scheme,
the other 11 components are determined.
Notice that one of the $A_{2}$-modes represents a ferromagnetic spin
arrangement, with the spins in the z-direction.  The in-plane (in the x and
y-direction) ferromagnetic spin arrangements are  
E-modes.
        
	In constructing the spin holograms (\ref{e:ialpha}), we take linear 
combinations of transformed emission patterns, $I(G^{-1}[\hat{\bf R}],
G^{-1}[\hat{\epsilon}];m)$.  If $\epsilon = \epsilon_{0} (= \hat{\bf z})$, then
since $G[\hat{\epsilon}_{0}] = \hat{\epsilon}_{0}$ for all elements $G$ of
$C_{4v}$, only a single photoelectron hologram, $I_{0}(\hat{\bf R})$, 
must be measured. For notational  
convenience we indicate in what follows the photon polarization by means
of a subscript.
With circularly polarized light incident along the 
z-direction, $\hat{\epsilon} = \hat{\epsilon}_{+}, \; \hat{\epsilon}_{-}$, 
the two intensity holograms, $I_{+}$ and $I_{-}$, suffice to construct
the projection holograms (\ref{e:ialpha}). 

	Representing the symmetry operation by the corresponding change
in the azimuthal angle dependence $\{$e.g. $I_{0}(s_{V}[\hat{\bf R}])$ is
represented by $I_{0}(\pi-\phi)$ etc...$\}$, we can write the spin holograms 
for the $A_{2}$, $B_{1}$ and $B_{2}$-modes as
\begin{eqnarray}
I^{A_{2}} (\hat{\epsilon}_{0} ;\hat{\bf R}) &=& 
\frac{1}{8} \; \{
I_{0}(\phi) + I_{0}(\phi+\pi/2) +  I_{0}(\phi-\pi/2) +  I_{0}(\phi+\pi)
\nonumber \\
&& -  I_{0}(\pi-\phi) - I_{0}(-\phi) -  I_{0}(\pi/2-\phi) - I_{0}(-\phi-\pi/2)
\} \; \; ,
\nonumber \\
I^{B_{1}} (\hat{\epsilon}_{0}; \hat{\bf R}) &=&
\frac{1}{8} \; \{
I_{0}(\phi) - I_{0}(\phi+\pi/2) -  I_{0}(\phi-\pi/2) +  I_{0}(\phi+\pi)
\nonumber \\
&& +  I_{0}(\pi-\phi) + I_{0}(-\phi) -  I_{0}(\pi/2-\phi) - I_{0}(-\phi-\pi/2)
\} \; \; ,
\nonumber \\
I^{B_{2}} (\hat{\epsilon}_{0}, \hat{\bf R}) &=&
\frac{1}{8} \; \{
I_{0}(\phi) - I_{0}(\phi+\pi/2) -  I_{0}(\phi-\pi/2) +  I_{0}(\phi+\pi)
\nonumber \\
&& -  I_{0}(\pi-\phi) - I_{0}(-\phi) +  I_{0}(\pi/2-\phi) + I_{0}(-\phi-\pi/2)
\} \; \; ,
\label{e:mod01}
\end{eqnarray}
where we suppressed the dependence on the polar angle, $\theta$, 
which is invariant under $C_{nv}$ symmetry elements.
For the one-dimensional $A$ and $B$-modes, the coefficients can
simply be read from the character table (table I).
For the two-dimensional $E$-representation, on the other hand,
we need to construct an actual representation.   The result is:
\begin{eqnarray}
I^{(E,1)} (\hat{\epsilon}; \hat{\bf R}) &=&
\frac{1}{4} \; \{
I_{0} (\phi) - I_{0} (\phi + \pi) + I_{0}(\pi-\phi) - I_{0}(-\phi) \} \; \; ,
\nonumber \\
I^{(E,2)} (\hat{\epsilon}, \hat{\bf R}) &=&
\frac{1}{4} \; \{
I_{0} (\phi) - I_{0} (\phi + \pi) - I_{0}(\pi-\phi) + I_{0}(-\phi) \} \; \; .
\label{e:mod02}
\end{eqnarray}
The analoguous spin holograms for the circularly polarized photons, 
are
\begin{eqnarray}
I^{A_{2}} (\hat{\epsilon}_{+};\hat{\bf R})
&=& \frac{1}{8} \; \{
I_{+}(\phi) + I_{+} (\phi+\pi/2) + I_{+}(\phi-\pi/2) + I_{+} (\phi+\pi)
\nonumber \\
&& - I_{-}(\pi-\phi) - I_{-}(-\phi) - I_{-}(\pi/2 - \phi) - I_{-}(-\phi-\pi/2)
\} \; \; ,
\nonumber \\
I^{B_{1}} (\hat{\epsilon}_{+};\hat{\bf R})
&=& \frac{1}{8} \; \{
I_{+}(\phi) - I_{+} (\phi+\pi/2) - I_{+}(\phi-\pi/2) + I_{+} (\phi+\pi)
\nonumber \\
&& + I_{-}(\pi-\phi) + I_{-}(-\phi) - I_{-}(\pi/2 - \phi) - I_{-}(-\phi-\pi/2)
\} \; \; ,
\nonumber \\
I^{B_{2}} (\hat{\epsilon}_{+};\hat{\bf R})
&=& \frac{1}{8} \; \{
I_{+}(\phi) - I_{+} (\phi+\pi/2) - I_{+}(\phi-\pi/2) + I_{+} (\phi+\pi)
\nonumber \\
&& - I_{-}(\pi-\phi) - I_{-}(-\phi) + I_{-}(\pi/2 - \phi) + I_{-}(-\phi-\pi/2)
\} \; \; . 
\nonumber \\
I^{(E,1)} (\hat{\epsilon}_{+};\hat{\bf R}) &=& 
\frac{1}{4} \; \{
I_{+}(\phi) - I_{+} (\phi+\pi) + I_{-}(\pi-\phi) - I_{-}(-\phi) \} \; \; ,
\nonumber \\
I^{(E,2)} (\hat{\epsilon}_{+};\hat{\bf R}) &=& \frac{1}{4} \; \{ 
I_{+}(\phi) - I_{+} (\phi+\pi) - I_{-}(\pi-\phi) + I_{-}(-\phi) \} \; \;  .
\label{e:mod2}
\end{eqnarray}

	Eqs. (\ref{e:mod01}), (\ref{e:mod02}) and (\ref{e:mod2}) 
constitute an analysis
of the measured hologram $I(\hat{\bf R}, \hat{\epsilon},m)$ into its six
components (adding the $A_{1}$ component to those given) appropriate to
$C_{4v}$-symmetry.  The advantage of this mode of analysis is that each
component is the hologram of a relatively simple spin configuration
${\bf s}_{(N)}^{(\alpha,i)}$, as given in Fig.(3) for each neighboring star of
atoms.  Determination of the components ${\bf s}_{(N)}^{(\alpha,i)}$  then
gives ${\bf s}_{N} (= \sum_{(\alpha,i)} {\bf s}_{(N)}^{(\alpha,i)})$.

\section{Extracting information from the spin holograms}

	How then, do we get the ${\bf s}^{(\alpha,i)}_{(N)}$ from the
hologram $I^{(\alpha,j)}$?  There are two main methods.  First we can
transform the hologram to give images of a sort of the individual atoms.
These `images' take the form shown in Eq.(\ref{e:imias}) below of standing 
spherical waves centered on the various atoms with strengths proportional
to ${\bf s}^{(\alpha,i)}_{(N)}$.  The advantage of this approach is that
it gives a 3D `picture' of the spin distribution on the neighboring atoms.
In fact, although these holographic images are of interest in particular
for obtaining a first qualitative determination of the spins, we think
that the most efficient and accurate spin determination will involve
iterative least square fitting of the direct hologram $I^{(\alpha,j)}
({\hat{\bf R}})$.

	We first discuss the nature of the holographic image.
The field of the holographic image of
a traditional (optical) spherical wave 
hologram $I(\hat{\bf R})$, formed by irradiating the negative
of the hologram with a spherical ingoing wave, can be shown \cite{Bart}
to be proportional to
\begin{equation}
{\cal I}({\bf r}) = \int I(\hat{\bf R}) \; \exp(ik \hat{\bf R} \cdot {\bf r}) 
\; d \Omega_{\hat{\bf R}} \; \; \; \; ,
\label{e:simtr}
\end{equation}
where $d \Omega_{\hat{\bf R}}$ denotes an infinitesimal solid angle of the
emission direction $\hat{\bf R}$.
In electron emission holography, transforms of the type of Eq.(\ref{e:simtr})
are called images, and as we shall see the atoms contributing to the
hologram appear in the image as centers of spherical waves.

	Consider the simple example of a
hologram formed by a spherically symmetric primary
wave, $\exp(ikr)/r$, 
scattered coherently with scattering amplitude $f$ by atoms i 
at positions ${\bf r}_{i}$. 
In the region $k|{\bf r}-{\bf r}_{i}| >> 1$, the wave scattered by atom i is
\begin{equation}
[\exp(ikr_{i})/r_{i}] \; f(\widehat{{\bf r}-{\bf r}_{i}};{\bf r}_{i}) \;
\exp(ik|{\bf r}-{\bf r}_{i}|)/|{\bf r}-{\bf r}_{i}| \; \; \; \; , 
\label{e:scatw}
\end{equation}
where 
$\widehat{{\bf r}-{\bf r}_{i}}$ denotes 
$({\bf r}-{\bf r}_{i})/|{\bf r}-{\bf r}_{i}|$.  The interference between
the primary and scattered waves then gives holographic fringes in
the emission intensity in the far-region, corresponding to
\begin{equation}
\sum_{i} \;
2 \; Re \{ [\exp(ikr_{i})/r_{i}] \; f(\hat{\bf R};\hat{\bf r}_{i}) \;
\exp(-ik \hat{\bf R} \cdot {\bf r}_{i}) \; \} \; \; .
\label{e:intt}
\end{equation}
Although the exchange scattering and the angular correlations expressed
by the interference matrix elements
complicates the expressions,
we can still recognize Eq.(\ref{e:intt}) in the projection 
hologram $I^{(\alpha,j)}$ of Eq.(\ref{e:ia}), replacing 
$f(\hat{\bf R};\hat{\bf r}_{i})$ by an `effective' 
scattering amplitude: $f \rightarrow f^{(eff)}$, where 
\begin{equation}
f^{(eff)}
(\hat{\bf R};\hat{\bf r}_{i}) = 
\langle \langle \hat{\bf R} | \sigma | \hat{\bf r}_{i} \rangle \rangle 
\cdot {\bf s}^{(\alpha,j)}_{i} \;
f_{s} (\hat{\bf R}; \hat{\bf r}_{i})  \; \; \; ,
\label{e:feff}
\end{equation}
and where ${\bf s}^{(\alpha,j)}_{i}$ 
is the spin on atom i in the ($\alpha ,j$)-projection 
of the ${\bf s}_{(N)}$-spin vector.
The interference terms (\ref{e:intt}) then contribute `imaging' terms
$\sum_{i} \; {\cal I}_{i}({\bf r})$ to the transform, where
\begin{eqnarray}
{\cal I}_{i} ({\bf r}) = 
[\exp(ikr_{i})/r_{i}] \; \int 
f^{(eff)}(\hat{\bf R};\hat{\bf r}_{i})
\exp(ik \hat{\bf R} \cdot [{\bf r} - {\bf r}_{i}]) \;
d \Omega_{\hat{\bf R}} \; \; \; \; .
\label{e:imi}
\end{eqnarray}

	We calculate the image by expanding the exponential factor in partial
waves, $\exp(ik\hat{\bf R}\cdot [ {\bf r}-{\bf r}_{i}])$ $ = 
4\pi \sum_{l,m} (i)^{l} \; j_{l} (k|{\bf r}-{\bf r}_{i}|)
Y_{lm} (\widehat{{\bf r}-{\bf r}_{i}}) Y^{*}_{lm}(\hat{\bf R})$,
where $j_{l}$ denotes the spherical Bessel functions of the second kind and
the $Y_{lm}$-functions are the spherical harmonics.  Similarly, $f^{(eff)}
(\hat{\bf R};\hat{\bf r}_{i}) = \sum_{LM} F_{LM} Y_{LM}(\hat{\bf R})$, and
assuming for the sake of the argument that the hologram is collected over the
full $4\pi$ sphere of emission directions, we obtain 
a closed expression for ${\cal I}_{i}$:
\begin{equation}
{\cal I}_{i} ({\bf r}) =
[\exp(ikr_{i})/r_{i}] \; 4 \pi \sum_{LM} F_{LM} \; (i)^{L} \; 
j_{L}(k|{\bf r}-{\bf r}_{i}|) \; Y_{LM}(\widehat{{\bf r}-{\bf r}_{i}})
\; \; \; , 
\label{e:imicl}
\end{equation}
where we made use of the orthonormality
of the spherical harmonics.
From Eq.(\ref{e:imicl}), we see that the holographic image is in fact
a sum of standing partial waves.
Notice that only the spherically symmetric (L=0) contribution in 
Eq.(\ref{e:imicl})
actually peaks at the position of the
scattering atom.  For nonzero values of L, the $j_{L}$ Bessel functions
reach their maximum value away from the origin \cite{foot1}.  
It is of some interest to note that
in the region $k|{\bf r}-{\bf r}_{i}| >> 1$,
using the asymptotic expansion of the Bessel functions,
$j_{l}(x) \sim \sin(x-l\pi/2)/x$, and with
 $(-1)^{L} Y_{LM}(\hat{\bf r}) = Y_{LM}(-\hat{\bf r})$, the image 
${\cal I}_{i}$ is a simple superposition of incoming and
outgoung waves: 
\begin{eqnarray}
{\cal I}_{i} ({\bf r}) &\sim& \; \frac{2\pi}{ik} \; \sum_{LM} \;
F_{LM} \; \{  Y_{LM}(\widehat{{\bf r}-{\bf r}_{i}}) \;
[\exp(ik|{\bf r}-{\bf r}_{i}|)
/|{\bf r}-{\bf r}_{i}| ] \nonumber \\
&& \; \; \; \; \; \; \; \; \; \; \; \; \; \; \; \; \;
- Y_{LM}(-[\widehat{{\bf r}-{\bf r}_{i}}])  \;
[ \exp(-ik|{\bf r}-{\bf r}_{i}|)
/|{\bf r}-{\bf r}_{i}| ] \}
\nonumber \\
&=& \frac{2\pi}{ik} \; \; \{ f^{(eff)} (\widehat{{\bf r}-{\bf r}_{i}}) ;
\hat{\bf r}_{i}) \; [ \exp(ik|{\bf r}-{\bf r}_{i}|)
/|{\bf r}-{\bf r}_{i}| ]
\nonumber \\
&& \; \; \; \; \; \; \; \; \; \;
-  f^{(eff)} (-[\widehat{{\bf r}-{\bf r}_{i}}]) ;
-\hat{\bf r}_{i}) \; [ \exp(-ik|{\bf r}-{\bf r}_{i}|)
/|{\bf r}-{\bf r}_{i}| ] \} \; \; .
\label{e:imias}
\end{eqnarray}

	Returning to the problem of determining the spins,
we assume that the scattering amplitudes $f_{s}$ in Eq.(\ref{e:feff})
can be calculated numerically 
and can be used in the analysis of the hologram.  
By analyzing the holographic image of the spin holograms and by determining
the particular mixture of partial waves
in the image (s,p,d etc...), one can then, in principle, find 
the spin vectors.  

	However, regarding the use of calculated scattering amplitudes,
we should remark that most numerical schemes 
assume spherical symmetry for the 
scatterer.  While this assumption is mostly justified for the spin
independent scattering amplitude $f_{0}$, it is usually not for the
spin dependent 
exchange scattering from the valence shell.
As a matter of fact, the exchange scattering amplitude 
$f_{s}$ is a quantity that is interesting in its own right,
as we discuss in the next section.
Note that the projection holograms give the possibility of determining
$f_{s}(\hat{\bf R},\hat{\bf r}_{i})$ (and not just $|f_{s}|^{2}$).
In many cases of interest, the spins are ordered in an arrangement with
known projection onto a particular ($\alpha ,j$)-spin mode, or it might be
possible to order the spins into such an arrangement, for example
by applying an external magnetic field (in the case of $C_{4v}$-symmetry
discussed above, one might be able to align the spins along the 
z-axis which is an $A_{2}$-mode, or along the x-axis, which
is an $(E,1)$-mode.  Under those circumstances the spin directions
of the ($\alpha,j$)-mode are known and only the exchange
scattering amplitude $f_{s}(\hat{\bf R},\hat{\bf r}_{i})$
needs to be determined in the expression for the projection
hologram,  which can be achieved by means of a least-square fit. 

	In fact, although holographic spin images are of interest, we think 
that the most efficient and accurate spin determinations will
involve fitting the hologram in a scheme that is similar to the
procedure to determine $f_{s}$, with 
the coefficients of the different spin modes of the same
($\alpha,j$)-representation as additional parameters to be determined.
One can devise an iterative scheme to determine $f_{s}$ and the spins:
Use the calculated $f_{s}$ to estimate 
the atomic spin vectors.
Then, use the obtained spins to refine the scattering
amplitude $f_{s}$ through a least-square fit.  
The resulting $f_{s}$ is then used
to obtain a better fit of the spins, and so on, until convergence is reached. 

\section{ Note on Rare Earths and Spin Holography}

        In our discussion in the preceeding sections we assumed that the
spin dependent part of the coherent scattering amplitude, see Eq.
(\ref{e:f}),
for an atom `j' was of the form
$F_{s} \sim {\bf \sigma} \cdot \langle {\bf S}_{j} \rangle \; f_{s} ({\bf k},
{\bf k}')$.  This is a good approximation, however, only when we can
assume that the spin density of atom j is of the form
${\bf s}_{j} ({\bf r}) = {\bf S}_{j} \rho_{j}
({\bf r})$; that is, that the orbital angular moments are `quenched'.
This is usually a good assumption for transition elements in crystals; 
but not for rare earth (or actinide) elements where there are strong
spin-orbit correlations.  In that case, we obtain (see appendix)
\begin{equation}
F_{s}({\bf k},{\bf k}') = {\bf \sigma} \cdot
\sum_{K = 0, M = - K}^{2l, K} A_{K} \;
T_{K M}^{\ast} (\hat{\bf k},\hat{\bf k}') \langle {\bf S}_{j}
{\cal Y}_{K M} ({\bf L}_{j}) \rangle \; .
\label{e:fgen}
\end{equation}
In the equation $F_{s}$ is the spin-dependent coherent exchange
contribution to the scattering amplitude (${\bf k} \rightarrow {\bf k}'$)
from a rare earth ion `j' where Russel-Saunders coupling is assumed to apply
with ${\bf S}_{j}$ and ${\bf L}_{j}$ the good spin and orbital angular
momentum of the ion. ${\cal Y}_{K M}({\bf L})$ is an irreducible tensor
constructed from the angular momentum operator ${\bf L}$, and
$T_{K M} (\hat{\bf k},\hat{\bf k}')$ is an irreducible tensor
depending on
$\hat{\bf k}$ and $\hat{\bf k}'$.  $A_{K}$ involves radial exchange
integrals as well as other coupling coefficients (see appendix).

        For the rare earths l=3 and Eq.(\ref{e:fgen}) gives $F_{s}$ in terms
of the scattering from the six multipole moment tensors of the
uncompensated spin distributions of the rare earth ion.

        If the spin and orbits are not coupled then
\begin{equation}
F_{s} = {\bf \sigma} \cdot \langle {\bf S}_{j} \rangle
\sum_{K , M} A_{K} T_{K M}^{\ast} (\hat{\bf k},\hat{\bf k}')
\langle {\cal Y}_{K M} ({\bf L}_{j}) \rangle \; ,
\label{e:fquen}
\end{equation}
and we have the quenched orbital result of Eq.(\ref{e:f}).

        In Eqs.(\ref{e:fgen}) and (\ref{e:fquen}), $K = 0$ is
the isotopic scattering term,
$T_{0} (\hat{\bf k},\hat{\bf k}') =
f_{0} (\hat{\bf k} \cdot \hat{\bf k}')$, $T_{1}$ the
vector term, $T_{1}(\hat{\bf k},\hat{\bf k}')
= \hat{\bf k} \; a_{11}(\hat{\bf k} \cdot
\hat{\bf k}') + \hat{\bf k}' \; a_{12} (\hat{\bf k} \cdot {\bf k}')
+ \hat{\bf k} \times
\hat{\bf k}' \; a_{13}(\hat{\bf k} \cdot \hat{\bf k}')$.
Similarly, $T_{20} (\hat{\bf k},
\hat{\bf k}') = (\hat{\bf k}_{z}^{2} - 1/3) \; a_{21}(\hat{\bf k}
\cdot \hat{\bf k}')
+ (\hat{\bf k}'^{2}_{z} -1/3) \; a_{22} (\hat{\bf k} \cdot \hat{\bf k}') +
(\hat{\bf k}_{z} \hat{\bf k}_{z}' -1/3 \hat{\bf k} \cdot \hat{\bf k}') \;
a_{23}(\hat{\bf k}
\cdot \hat{\bf k}') , \cdots $.

        In the hypothetical case of no spin-orbit coupling then the
spin $\langle
{\bf S}_{j} \rangle$ would be free to vary independent of the anisotropic charge
distribution and would correspond to our previously discussed theory.  In fact,
however, magnetic studies show \cite{Mac} that $J = |{\bf L} + {\bf S}|$ retains
its Hunds rule ground state value, $ J = L + S $, for the second half shell, and
$ J = L - S $  for the first half shell in the crystal
(except $Eu^{3+}$ and $Ce^{3+}$)
to a good approximation.

        In that case Eq.(\ref{e:fgen}) becomes
\begin{equation}
F_{s} = \sum_{K'=K-1,K=0,M'}^{K+1,2l}
 A_{K} B_{K K'} (\sigma \otimes T_{K}
(\hat{\bf k}, \hat{\bf k}'))^{\ast}_{K'M'}
\langle {\cal Y}_{K' M'} ({\bf J}) \rangle \; \; \; ,
\label{e:fhund}
\end{equation}
where $B_{K K'}$ is another coupling coefficient (see Appendix) and
\begin{equation}
(\sigma \otimes T_{K}(\hat{\bf k},\hat{\bf k}'))_{K'M'}
\equiv
\sum_{\mu,\nu}  C_{\mu \; \nu \; M'}^{1 \; K \; K'}
\sigma_{\mu} T_{K \nu} (\hat{\bf k},\hat{\bf k}') \; .
\label{e:t}
\end{equation}

	\underline{Discussion.}
Now, instead of just $\langle {\bf S} \rangle$ or $\langle {\bf J} \rangle $ to
represent the spin
dependent coherent exchange scattering, there are seven, 2l+1, multipole moment
distributions $\langle {\cal Y}_{K M} ({\bf J}) \rangle$
giving $(2l+1)^{2} = 49$
parameters determining the scattering.
$Gd^{3+}$ is uniquely simple because it is
spherically symmetric so that in Eq.(\ref{e:fgen})
$\langle {\cal Y}_{K M} ({\bf L}) \rangle =
7 \delta_{K 0}$ and thus $F_{s} \sim \sigma \cdot \langle {\bf J} \rangle
f_{0}(\hat{\bf k} \cdot \hat{\bf k}')$.
For the other rare earths however, and for $E_{k,k'}
\simeq 100 \; {\rm eV}$, tensor components up to $K \simeq 3,4$ make important
contributions to $F_{s}$.

        The question now arises, are there good reasons to measure and
analyze `spin
holograms' of rare earth ions, given their evident complexity?
We think there is
because of the relation of such scattering measurements to the RKKY mechanism
responsible for the rare earth ion interactions in solids.
The interaction arises from
the exchange scattering of conduction electrons with 4f shell
electrons of an ion
inducing a stationary spin polarization wave in the surrounding conduction band.
The role of the anisotropic, higher moment terms, Eq.(40), in $F_{s}$ on the
RKKY-interaction have not been investigated as thoroughly as their
importance warrants.  Frederick Specht \cite{Specht} in an early investigation
retained only terms up to 
$K =2$ and found quite large effects.  Thus for two nearest neighbor $Tb$
ions whose moments were aligned along the inter-ionic direction he
obtained $E(0)
= -6.6 k_{B}$, while if the moments are perpendicular to that direction
he obtained
$E(\pi/2) = - 8.1 k_{B}$, 
a $20 \%$ variation; while for Tm ions the numbers were 
$E(0) =-1.2  k_{B},  E(\pi/2) = -.7 k_{B}$, a $50 \%$ variation.
(Note that for the cigar shaped spin distribution of Tb $|E(\pi/2)| > |E(0)|$,
while the opposite held for the pancake shaped Tm spin distributions).

        We think that spin holographic experiments on a ferromagnetic surface
layer of terbium, for example, as a function of temperature, magnetic field,
and $k$ of the photoelectron, revealing the effects of the moments of various
orders, would be very interesting.

	\underline{Measurement Theory.}
An interesting question is what measurement must be made to
determine the quantum state of a system \cite{Gayle}.  Consider a system with
angular momentum $J$.  In general its state is determined by its density
matrix $\rho_{M,M'}$.  What measurements can be made to determine the
$(2J+1)^{2}-1 = 4J(J+1)$ real numbers required to specify
$\rho$?  In reference \cite{Gayle} we outlined one
method using `Feynman filters'; but we also mentioned that Fano \cite{Fano} had
shown that measurement of the $4J(J+1)$ expected multipole moments
$\langle {\cal Y}_{K M} ({\bf J})
\rangle, K =1,....,2J$ also would suffice.
In our work here we have seen that the
$\langle {\cal Y}_{K M} ({\bf J}) \rangle$'s could be determined
(in principle) for $K \leq 2l+1$, or $K \leq 2L+1$ by means of
coherent electron scattering, the higher order multipoles vanish.  Thus for
Gd only the first order moment, $\langle
{\bf J} \rangle$ could be so determined, while for the other rare earths
much more state information can be determined, but not enough to fully
determine the state.

\section{Conclusions}

        The main message of this paper is that we can construct holograms
of the atomic spins in the near-neighborhood of a photoelectron emitting atom
from the angularly resolved photoelectron emission intensities.
For the important case of photoelectrons emitted by source atoms
with $C_{nv}$-environment, we show how these spin holograms
can be constructed using symmetry.  Furthermore, we discuss
schemes to holographically image the
spins and to extract accurate spin information from the spin holograms.

The photons used to emit the electrons that give the intensity holograms
for the purpose of spin holography can be, ${\it but \;
do \; not \; have \; to \; be}$, circularly polarized.  Incident
linearly polarized,
or indeed even ${\it unpolarized}$ light would also serve the purpose.
This statement could appear somewhat puzzling because it
seems to imply that it is
unnecessary to polarize the primary photoelectron waves in order to
probe the spins of the nearby scattering atoms.  We show that the surprising
ability of unpolarized light to record spin information in the photoelectron
intensity is a
consequence of the spin-orbit correlations in the initial inner-core
electron states from which the photoelectron is emitted.  The spin-orbit
correlations
give a finite interference contribution to the spin
of electrons emitted in different directions.
It is the interference of
the photoelectron emitted in the $\hat{\bf R}$-direction,
and that emitted in the direction of atom i and then scattered into the
$\hat{\bf R}$-direction, that gives the holographic fringes in the
photoelectron intensity.
By virtue of
exchange scattering the holographic fringes are then sensitive
to the spin of the scattering atom, with contributions proportional to
the scalar product of the atomic spin and the interference contribution
to the photoelectron spin,
$\langle \langle \hat{\bf R} | \sigma |  \hat{\bf r}_{i} \rangle
\rangle$.  Note that it is not, as one might have surmised,
the spin of the photoelectron wave emitted in the ${\bf r}_{i}$-direction,
$\langle \langle \hat{\bf r}_{i} | \sigma | \hat{\bf r}_{i} \rangle \rangle$,
that is of importance.  Indeed, under some conditions, this primary wave
spin polarization can vanish,  implying unpolarized electrons impinging on the
scattering atom, while  $\langle \langle \hat{\bf R} | \sigma | \hat{\bf r}_{i}
\rangle \rangle \neq 0$, implying a spin dependent photoelectron intensity
hologram. In the paper, we calculate the spin intereference matrix
element for photoemission from a $p_{1/2}$
shell, by absorbing photons of arbitrary polarization.

        Furthermore, to obtain an actual spin hologram from the
photoelectron intensity, it is necessary to `separate
out' the spin dependent holographic fringes, which typically contribute
5 $\%$ or less, from the rest of the hologram.
In this paper, we point out how this separation
can be obtained using symmetry if the source atom is in a site of 
$C_{nv}$-symmetry.
Specifically, we show that a group theoretical projection of the
emission intensity onto the irreducible ($\alpha,j$)-representation gives a
hologram of the projection of the spin arrangement of
the nearby atoms onto the ($\alpha,j$)-spin mode.
This way, one can obtain almost all components of the spin vectors
(all but one for the n-atom star and all but three for the 2n-atom star,
as explained in the text).

        The advantage of measuring emission intensities, rather than
having to spin analyze the emitted electrons, is considerable: one does
not lose the usual factor of $10^{4}$ in intensity due to the
low efficiency of the electron polarimeters.  Consequently, we expect that
the proposed experiments are quite feasible at a synchrotron source and
it is possible that the data have already been obtained and only need to be
analyzed.

        To conclude, we repeat that spin holography is a surface probe
which probes the average short-range magnetic environment of the photoelectron
source atom.  By using atoms adsorbed on a surface as sources, spin holography
can reveal the influence of these adsorbates on the spins of the neighboring
substrate atoms.  It can then provide detailed information on many interesting
systems, such as adsorbed oxygen atoms quenching the magnetism of a nickel
surface, or iron atoms inducing moments on  neighboring palladium
substrate atoms (superparamagnetism).  Alternatively, substrate atoms can be
used as sources to image the spins of objects adsorbed on the surface, such
as small molecules. Finally, spin holography should be useful for surface
studies of magnetic systems with large and complicated unit cells of
N spins, such as garnets.  (Spin holography gives information about the
N spins as a function of the N positions, whereas the usual diffraction
techniques give the $N^{2}$ spin-spin correlation as functions of the
$N^{2}$ relative positions).

\section*{Acknowledgments}

Part of the work was carried out with support from National Science
Foundation, Grant No DMR-9013058.
The work of E.T. is supported by the National Science Foundation
through a grant for the
Institute for Atomic and Molecular Physics at Harvard University
and Smithsonian Astrophysical Observatory.

\newpage

\section*{Appendix}

        We shall sketch the derivation of Eqs.(38-40).

        If ${\bf s}_{c}$ is the spin of the positive energy `free' electron
and ${\bf s}_{i}$ that of a bound electron then the spin exchange operator
is $1/2 + 2 {\bf s}_{c} \cdot
{\bf s}_{i}$, and the exchange scattering amplitude for free electron
momentum ${\bf k}$ and bound electron in orbital $|u_{lm}\rangle $
going to a free electron momentum ${\bf k}'$ and bound electron
$\rightarrow |u_{lm'}\rangle $ is
\begin{equation}
f_{s} = (1/2 + 2 {\bf s}_{c} \cdot {\bf s}_{i} )
\; {\cal S} ({\bf k},u_{lm};{\bf k}',u_{lm'}) \; \; ,
\label{e:a1}
\end{equation}
where
\begin{equation}
{\cal S} = - \int \exp[-i{\bf k}' \cdot {\bf r}_{2}] \; u^{\ast}_{lm'}
({\bf r}_{1})
\frac{e^{2}}{r_{12}} u_{lm}({\bf r}_{2}) \; \exp[i{\bf k} \cdot {\bf r}_{1}] \;
d^{3} r_{1} d^{3} r_{2} \; \; ,
\label{e:a2}
\end{equation}
is the exchange integral.

        Expanding the integral in partial waves we get
\begin{equation}
{\cal S} = \sum_{p\; p' \; \lambda} E(l,p,k;l,p',k';\lambda) 
\sum_{q\; q' \; \mu}
\langle l,m' | Y_{pq} | \lambda \mu \rangle \;
\langle \lambda \mu | Y_{p'q'} | lm \rangle
Y^{\ast}_{p'q'}(\hat{\bf k}') Y^{\ast}_{pq}(\hat{\bf k}) \; \; ,
\label{e:a3}
\end{equation}
where E involves radial overlap integrals
\begin{eqnarray}
&& E(l,p,k;l,p',k';\lambda) = 
\nonumber \\
&& -(i)^{p-p'} \frac{(4\pi)^{3} e^{2}}{(2p+1)(2p'+1)(2\lambda+1)}
\int j_{p'}(k'r_{2}) g_{l}(r_{1}) \frac{r^{\lambda}_{<}}{r^{\lambda+1}_{>}}
g_{l}(r_{2}) j_{p}(kr_{1}) r_{1}^{2} r_{2}^{2} dr_{1} dr_{2} \; ,
\label{e:a4}
\end{eqnarray}
with $j_{p}$ the spherical bessel function and $u_{lm}({\bf r}) = g_{l} (r)
Y_{lm}(\hat{\bf r})$.

        Now making use of the Racah-Wigner techniques (\cite{Racah}) we obtain
\begin{eqnarray}
&&{\cal S} = \sum_{p\; p' \; \lambda} \langle l || Y_{p} || \lambda \rangle
\langle \lambda || Y_{p'} || l \rangle \; \sqrt{2\lambda +1}
\; E(l,p,k;l,p',k';\lambda)
\nonumber \\
&& \; \; \; \; \; \; \; \; \cdot \; \sum_{K M} (-)^{K} \sqrt{2K+1}
\left\{ \matrix{l&p&\lambda \cr p'&l&K \cr } \right\}
\left[ \left( Y_{p'} (\hat{k}') \otimes Y_{p}(\hat{k})
\right)^{K}_{M} \right]^{\ast} C^{l \; \; K \; \; l}_{m \; M \; m'} \;  \; .
\label{e:a5}
\end{eqnarray}
We now replace the vector addition coefficient in (\ref{e:a5}) 
in accordance with
$C^{l \; \; K \; \; l}_{m \; M \; m'} =
\langle l m' | {\cal Y}_{K M} ({\bf l}_{i})
|l m \rangle / \langle l || {\cal Y}_{K} || l \rangle $,
where ${\cal Y}_{K
M} ({\bf l})$ is an irreducible tensor of rank $K$ constructed from the angular momentum operators ${\bf l}$.

        The exchange scattering operator is now obtained by multiplying
${\cal S}$ by the spin exchange operator and summing over the n electrons
in the $4f$ shell:
\begin{eqnarray}
F_{x} &=& \sum_{i}^{n} f_{s}(i)
\nonumber \\
&=& \sum_{K = 0, M = -K}^{2l, K}
A_{K} T^{\ast}_{K M} (\hat{\bf k},\hat{\bf k}')
\langle (1/2 + 2 {\bf s}_{c} \cdot
\frac{{\bf S}}{\nu} ) {\cal Y}_{K M} ({\bf L})
\rangle \;  \; ,
\label{e:a6}
\end{eqnarray}
where ${\bf L}$, ${\bf S}$ are the operators for the total orbital angular
momentum and spin of the $4f$ electrons (we assume Russel-Saunders coupling
with good L and good S).  Also, we have supposed the Russel-Saunders
ground state values of $L$ and $S$ and we made the substitution 
\begin{eqnarray}
&& \sum_{i=1}^{n} (1/2+2 {\bf s}_{c} \cdot
{\bf s}_{i}) \; {\cal Y}_{K M}
({\bf l}_{i}) 
\nonumber \\
&& \; \; \; \; \; \; \; \; \; \; \; \; \; \; \; \; \;  
= \pm  \frac{ \langle l L || \sum_{i=1}^{\nu} {\cal Y}_{K}
({\bf l}_{i}) || l L \rangle }
{ \langle L || {\cal Y}_{K} ({\bf L}) || L \rangle }
\; \langle (1/2 + 2 {\bf s}_{c} \cdot \frac{{\bf S}}{\nu} )
{\cal Y}_{K M} ({\bf L}) \rangle \; \; ,
\label{e:a7}
\end{eqnarray}
where $\nu = n$ for less than half filled shell, $\nu = N - n$ for
more than half filled.  For more than half filled shell we must
choose the minus sign in (\ref{e:a6}) and then add
$\frac{N}{2} \delta_{K 0} $ to the spin independent term (i.e. rather than
$-(\frac{\nu}{2} + 2 {\bf s}_{c} \cdot {\bf S})$ for the
$K = 0$ term we get
$(\frac{N-\nu}{2} - 2 {\bf s}_{c} \cdot {\bf S})$ for that term).  Thus finally
\begin{eqnarray}
F_{x} = && \eta \sum_{p,p',\lambda}  E(l,p,k;l,p',k';\lambda)
\langle l || Y_{p} | \lambda  \rangle \langle \lambda ||
Y_{p'} || l \rangle \sqrt{2\lambda
+1}
\nonumber \\
&& \sum_{K M} (-)^{K}
\sqrt{2K +1} 
\left\{ \matrix{l&p&\lambda \cr p'&l&K \cr } \right\}
\left[ Y_{p'} (\hat{\bf k}') \otimes
Y_{p} (\hat{\bf k}) \right]^{\ast}_{K M}
\nonumber \\
&& \; \frac{\langle l L || \sum_{i=1}^{\nu}
{\cal Y}_{K} ({\bf l}_{i}) || l L \rangle }
{ \langle L || {\cal Y}_{K} || L \rangle }
\; (1/2 + 2 {\bf s}_{c} \cdot \frac{{\bf S}}{\nu} )
{\cal Y}_{K M} ({\bf L}) \; \; ,
\label{e:a8}
\end{eqnarray}
where $\eta$ is (+/-) for (less than/more than) half filled shell systems.
The meaning of $A_{K} T^{\ast}_{K M}
(\hat{\bf k},\hat{\bf k}')$ in Eq.(\ref{e:a6}) can then be obtained by
compairing with Eq.(\ref{e:a8}).

        Finally we suppose that $J$ is also a good quantum number
(with $J = L - S$ for the first half and $J = L + S$ for the second
half rare earths).  Then within the manifold of good $L, S$ and $J$
\begin{equation}
{\cal Y}_{K M} ({\bf L}) = (-)^{L+S+J+K} \sqrt{(2L+1)(2J+1)}
\left\{ \matrix{J&S&L \cr L&K&J \cr } \right\}
\frac{\langle L || {\cal Y}_{K} ({\bf L}) || L \rangle }{\langle J ||
{\cal Y}_{K} ({\bf J}) || J \rangle } \;
{\cal Y}_{K M}({\bf J})  \; ,
\label{e:a9}
\end{equation}
and this may be substituted into the spin independent term in 
Eq.(\ref{e:a6}).  Similarly,
\begin{eqnarray}
&& \sum_{M} T^{\ast}_{K M} (\hat{\bf k},\hat{\bf k}') \; {\bf \sigma}
\cdot {\bf S} \; {\cal Y}_{K M} ({\bf L})
\nonumber \\
&& \; \; \; \; \; = \sum_{K' \nu} \left[ \sigma^{(1)} \otimes
 T_{K} \right]^{K' \ast}_{\nu}
\cdot \left[ {\bf S}^{(1)} \otimes
{\cal Y}_{K}({\bf L}) \right]^{K'}_{\nu}
\nonumber \\
&& \; \; \; \; \; =  \sum_{K' \nu} \left[ \sigma^{(1)}
\otimes T_{K} (\hat{\bf k},
\hat{\bf k}') \right] ^{K'}_{\nu} 
\left[ \matrix{S&L&J \cr 1&K&K' \cr S&L&J \cr } \right]
\langle S || S || S \rangle \langle L ||
{\cal Y}_{K}({\bf L}) || L \rangle
\frac{{\cal Y}_{K' \nu}({\bf J})}
{\langle J || {\cal Y}_{K'} ({\bf J}) || J
\rangle} \; \; ,
\label{e:a10}
\end{eqnarray}
and substituting (\ref{e:a10}) into Eq.(\ref{e:a6}) gives 
Eq.(\ref{e:fhund}) of the text.

\newpage

\begin{table}
\caption{Character table of the group $C_{4v}$}
\label{tab1}
\begin{tabular}{c c c c c c}
\multicolumn{1}{c}{$C_{4v}$}&
\multicolumn{1}{c}{$E$}&
\multicolumn{1}{c}{$2C_{4}$}&
\multicolumn{1}{c}{$C_{2}$}&
\multicolumn{1}{c}{$2\sigma_{v}$}&
\multicolumn{1}{c}{$2\sigma_{d}$}\\
\tableline
$A_{1}$&1&1&1&1&1 \\
$A_{2}$&1&1&1&-1&-1 \\
$B_{1}$&1&-1&1&1&-1 \\
$B_{2}$&1&-1&1&-1&1 \\
$E$&2&0&-2&0&0 \\
\end{tabular}
\end{table}

\newpage

\section*{Figure Captions}

\noindent
\underline{Figure 1:} Reference frame used in calculating the spinors
of the electrons emitted from a $p_{1/2}$-shell (section II).  The
photon polarization vector, $\hat{\epsilon}$, is parallel to the 
$\hat{\bf z}$-direction.

\noindent
\underline{Figure 2:} 4-atom environment of electron emitting atom (denoted
by the shaded sphere).  The symmetry-group of the environment is $C_{4v}$, with
the $C_{4}$-axis (= $\hat{\bf z}$-axis) perpendicular to the plane 
of the four atoms.

\noindent
\underline{Figure 3:} 
The group theoretical modes of the spin vectors of the $C_{4v}$-environment
shown in Fig.2.  The spins are shown in the xy-plane, $\odot$ 
indicates an up-spin (positive $\hat{\bf z}$-direction), and $\otimes$ 
indicates a `down'-spin (negative $\hat{\bf z}$-direction).


\begin{thebibliography}{99}
\bibitem{foot1}  The higher the value
of L, the further out this maximum is reached,  so that the anisotropy of the
scattered wave `degrades' the image for the purposes of determining the
positions of the scattering atoms.  For the low energy electrons ($ k \sim
1 \AA^{-1}$) this can lead to overlapping images (since the distance between
atoms is typically a few $\AA$).
In ordinary electron emission holography, problems of this kind are
overcome using an imaging transform with a kernel that `corrects' for
the anisotropy (for example by dividing the integrand
in the holographic transform by the scattering
amplitude), and by adding the holograms collected at different emission
energies with a `phase' factor $\sim \exp(-ikr_{i})$ that `phase locks'
onto the interference term with the wave scattered by atom $i$ (canceling
the phase of
the $[\exp(ikr_{i})/r_{i}]$ factor in (\ref{e:imicl})).
\end{thebibliography}
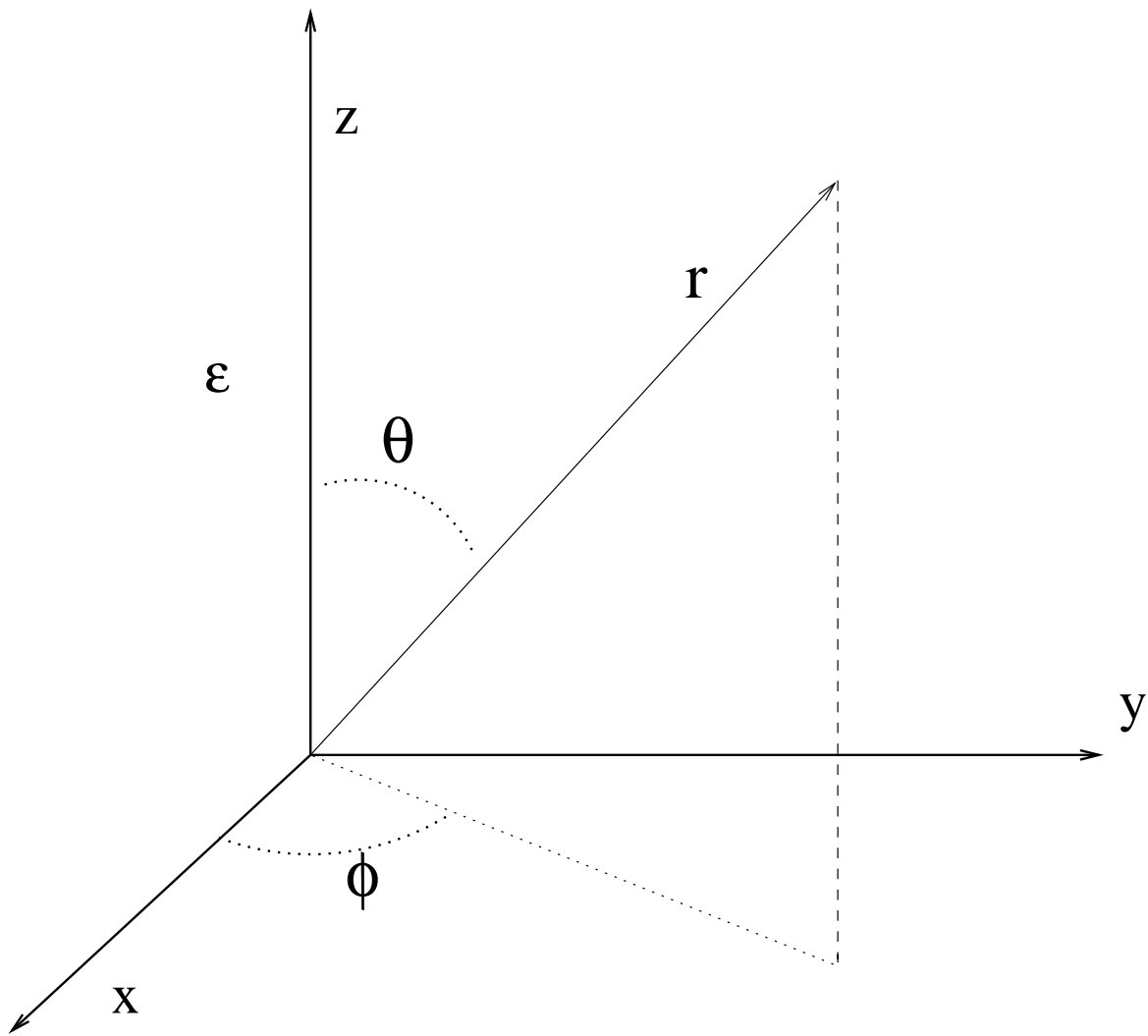
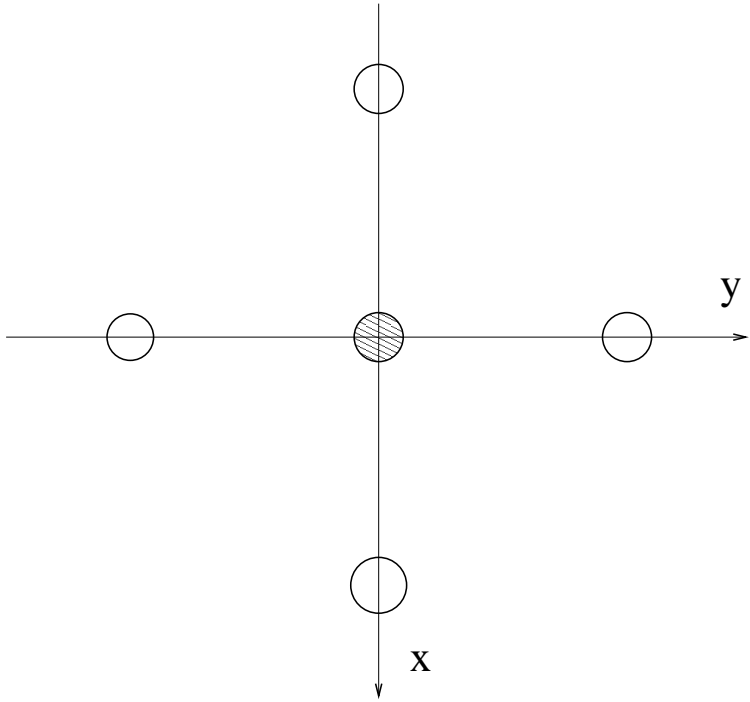
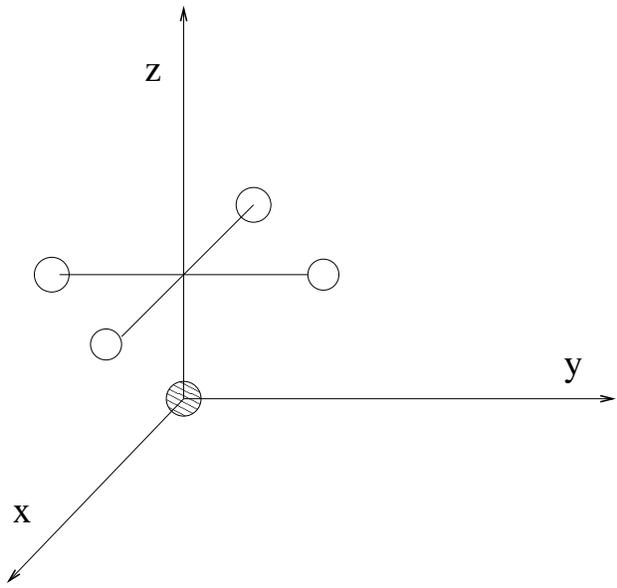
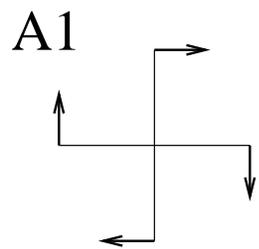
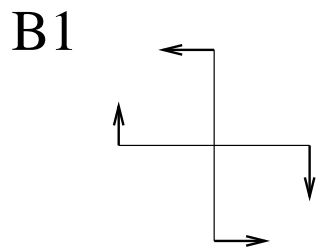
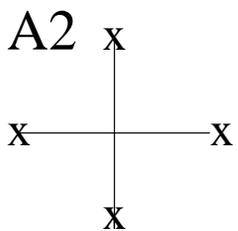
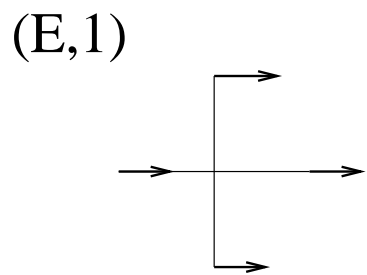
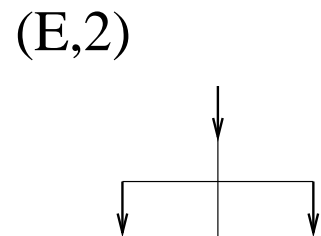
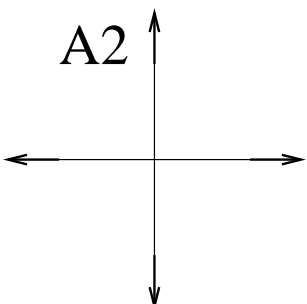
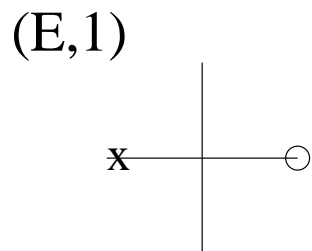
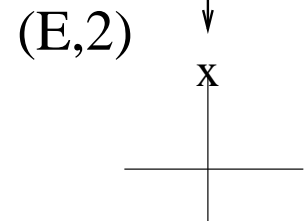
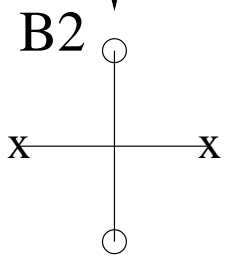
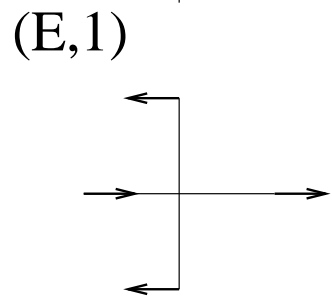
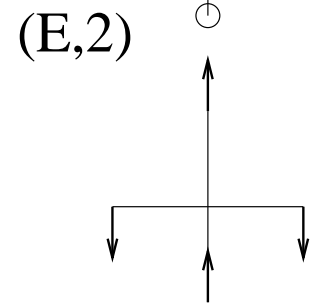
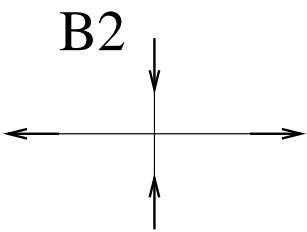
\end{document}